\documentclass[aps,prb,superscriptaddress,twocolumn]{revtex4-1}
\usepackage{amsmath,amssymb,amsthm,amsfonts}  
\usepackage[utf8]{inputenc}  
\usepackage[T1]{fontenc}
\usepackage[main=english,french]{babel}  
\usepackage[a4paper,vmargin=2.2cm,hmargin=2.0cm]{geometry}
\usepackage{mathrsfs}    
\usepackage{siunitx}     
\usepackage{braket}
\usepackage{cancel}
\usepackage{verbatim}     
\usepackage{graphicx}    
\usepackage{float}       
\usepackage{caption}     
\usepackage{hyperref}
\usepackage{tabularx}
\usepackage{multirow}
\usepackage{colortbl}
\usepackage{subfigure}
\usepackage{enumitem}
\setdescription{labelsep=\linewidth,leftmargin=0pt}
\usepackage[dvipsnames]{xcolor}
\usepackage{soul}
\usepackage{mathtools}

\usepackage{physics} 
\usepackage{bbm}

\newcommand{\dis}{{W_{0}}}

\begin{document}

\title{Excitons in Large Disordered Boron-Nitride Layer \\ using Linear-Scaling Bethe-Salpeter Simulations}

\author{Thomas Galvani}
\email{thomas.galvani@icn2.cat}
\affiliation{Catalan Institute of Nanoscience and Nanotechnology (ICN2), CSIC and BIST, Campus UAB, Bellaterra, 08193 Barcelona, Spain}

\author{Lorenzo Sponza}
\affiliation{Universit\'e Paris-Saclay, ONERA, CNRS, Laboratoire d'\'etude des microstructures (LEM), 92322 Ch\^atillon, France}

\author{Hakim Amara}
\affiliation{Universit\'e Paris-Saclay, ONERA, CNRS, Laboratoire d'\'etude des microstructures (LEM), 92322 Ch\^atillon, France}
\affiliation{Universit\'{e} Paris Cit\'{e}, Laboratoire Mat\'{e}riaux et Ph\'{e}nom\`{e}nes Quantiques, CNRS, F-75013 Paris, France}

\author{Sylvain Latil}
\affiliation{Universit\'e Paris-Saclay, CEA, CNRS, SPEC, 91191 Gif-sur-Yvette, France}

\author{Stephan Roche}
\affiliation{Catalan Institute of Nanoscience and Nanotechnology (ICN2), CSIC and BIST, Campus UAB, Bellaterra, 08193 Barcelona, Spain}
\affiliation{ICREA Institucio Catalana de Recerca i Estudis Avancats, 08010 Barcelona, Spain}

\begin{abstract}
We introduce a real-space, linear-scaling Bethe-Salpeter framework that enables excitonic spectroscopy in large and possibly disordered boron-nitride-derived systems. Thanks to the use of a sublattice-resolved perturbative decoupling that maps localized electron-hole pairs onto a sparse tight-binding model, we implement the Kernel Polynomial Method to compute absorption spectra with O(N) cost. To illustrate the capabilities of our method, we apply it to Anderson-disordered monolayer hexagonal boron nitride with up to $10^{5}$ orbitals. The method reveals a disorder-induced asymmetric broadening of bright excitons, a crossover from quadratic to linear redshift of the main absorption peak, and Anderson localization of the exciton center of mass. This approach extends excitonic calculations beyond the reach of conventional \textit{ab initio} Green's function methods (GW approximation and Bethe-Salpeter equation), opening optical spectroscopy to large-scale, disordered, moiré, quasicrystalline, and structurally complex quantum materials.
\end{abstract}


\maketitle

Excitons, as Coulomb-bound electron-hole pairs, dominate the optical spectra of many low-dimensional semiconductors and insulators, where quantum confinement and reduced dielectric screening enhance binding energies and oscillator strengths.\cite{Wannier1937,Elliott1957,Rytova1967,Keldysh1979,Cudazzo2011} The standard predictive route is many-body perturbation theory, in which quasiparticle corrections, often at the GW level, are combined with the Bethe-Salpeter equation (BSE)\cite{Salpeter1951} for the interacting electron-hole polarization.\cite{Hanke1980,Hybertsen1986,Rohlfing2000,NEW_Onida2002} This framework has reached a high level of accuracy for crystalline materials, including hexagonal boron nitride (hBN), whose strongly bound, short-radius excitons make it a prototypical system in which atomistic details matter.\cite{Arnaud2006,Galvani2016,Paleari2018,Sponza2018,Cudazzo2016}

The same excitonic physics becomes much harder to access in the systems now central to van der Waals materials research: disordered monolayers, moiré superlattices, large-angle quasicrystalline approximants, functionalised nano-objects and complex heterostructures. In such systems, optical spectra are controlled by the interplay between electron-hole attraction, local registry, and the breaking of translational symmetry.
Recent work on moir{\'e} excitons  shows that moderate disorder and finite temperature can suppress or reshape the very signatures used to identify correlated moir{\'e} exciton physics.\cite{Kim2026} The challenge is therefore not only to account for the excitonic physics, but to do so in atomistically large structures where symmetry cannot be relied upon.\cite{Stohler2026, Nalabothula2026}
If this can be achieved at a reasonable compromise between accuracy and computational workload for quasiparticle corrections and for the electron-hole interaction,~\cite{Rytova1967,Keldysh1979,Cudazzo2011,Trolle2017,Andersen2015,Gjerding2020, NEW_Ninhos2026, Kirichenko2019, Kirichenko2021}
solving the BSE with atomic resolution with light computational schemes is still the main hurdle. 
Several developments have addressed this bottleneck by adopting different strategies including Wannier or tight-binding local orbitals,\cite{UriaAlvarez2024,Dias2023, Maity2025} projection techniques,\cite{Naik2022,Li2024_imaging} or by avoiding the full diagonalization of the Hamiltonian, \cite{Schmidt2003,Fuchs2008} as well as recently proposed tensor-network approaches.\cite{Moustaj2026}

A particularly relevant advance is the real-space Wannier-function formulation of Merkel and Ortmann, which yields linear-scaling optical spectra once a sparse exciton Hamiltonian is constructed from maximally localized Wannier functions.\cite{Merkel2024} However, an efficient route is still missing for atomistic excitons in large disordered or quasicrystalline cells where  explicit eigenstate constructions, and thus Wannier or pristine-cell projections, are not possible.

In this Letter, we provide such a route for hBN-derived systems. Starting from a sparse tight-binding Hamiltonian with well-separated subspaces, we perturbatively construct effective electron and hole single particle Hamiltonians and map the BSE onto a sparse real-space Hamiltonian for localized electron-hole pairs. Truncating away the pairs with electron-hole separation above a fixed cutoff radius yields a Hilbert space that scales linearly with the number of atoms, while the Kernel Polynomial Method (KPM) evaluates the imaginary part of the dielectric function without diagonalization.\cite{Weisse2006,Fan2021} The method works in a perturbative framework, but gives access to regimes unfeasible to standard reciprocal-space calculations as it permits to attain up to millions of atoms yet using very moderate computational resources. As it enables calculations in extremely large supercells, it also makes possible novel kinds of analysis, like statistical investigations. Here, we illustrate its potential by studying an Anderson-disordered monolayer hBN with up to $10^{5}$ orbitals and show how disorder produces asymmetric broadening, redshifts the bright excitonic peak, activates a continuum of optical oscillator strength, and localizes the exciton center of mass while keeping the electron-hole pair compact. This approach thus introduces a scalable atomistic framework for exploring excitonic physics in disordered and large-unit-cell two-dimensional insulators.

Following prior works \cite{Galvani2016, Paleari2018, Sponza2018} we consider a system whose single particle electronic structure is described by a tight-binding Hamiltonian separable onto two sublattices \(\Lambda_{\mathrm{B}}\) and \(\Lambda_{\mathrm{N}}\):
\begin{equation*}
    \hat{H}_{el} =
    \begin{pmatrix}
    \Delta \mathbbm{1}_\mathrm{B} & 0 \\
    0 & -\Delta \mathbbm{1}_{\mathrm{N}}
    \end{pmatrix}
    +
    \begin{pmatrix}
    \hat{h}_\mathrm{BB} & 0 \\
    0 & \hat{h}_\mathrm{NN}
    \end{pmatrix}
    +
    \begin{pmatrix}
    0 & \hat{h}_\mathrm{BN} \\
    \hat{h}_\mathrm{NB} & 0
    \end{pmatrix}
\end{equation*}
here in a basis constructed as the concatenation 
$\mathcal{B}_{el}=\{\ket{\mathrm{B},\vb{n}}\}_{\vb{n}\in\Lambda_\mathrm{B}} \oplus \{\ket{\mathrm{N},\vb{p}}\}_{\vb{p}\in\Lambda_\mathrm{N}}$, with the \(\ket{\mathrm{B},\vb{n}}\) and \(\ket{\mathrm{N},\vb{p}}\) being atomic orbitals centered on the corresponding sublattice positions.
Here 
\(2\Delta>0\) is the main energy separation between the high (B) and low (N) energy states. The block matrices $\hat{h}_\mathrm{BB}$ and $\hat{h}_{\mathrm{NN}}$ correspond to the variations 
in onsites values and the hoppings inside the subspaces, 
while $\hat{h}_{\mathrm{BN}}$ and $\hat{h}_{\mathrm{NB}}$ describe their coupling, which is mediated by B-N hopping integrals. We then treat these couplings through second order perturbation theory to construct approximate electron and hole Hamiltonians
\begin{align}
\label{eq:Pert_Hams}
\begin{split}
\hat{H}_e &\approx \phantom{-}\Delta \mathbbm{1}_\mathrm{B} + \hat{h}_{\mathrm{BB}} + \frac{1}{2\Delta}\hat{h}_{\mathrm{BN}}\hat{h}_{\mathrm{NB}} \\ 
\hat{H}_h &\approx -\Delta \mathbbm{1}_{\mathrm{N}} + \hat{h}_{\mathrm{NN}} - \frac{1}{2\Delta}\hat{h}_{\mathrm{NB}}\hat{h}_{\mathrm{BN}}
\end{split}
\end{align}
respectively, and approximate the electron (resp. hole) subspace as being spanned by Boron (resp. Nitrogen) orbitals belonging to the \(\Lambda_\mathrm{B}\) (resp. \(\Lambda_\mathrm{N}\)) sublattice. Within this picture, the Bethe-Salpeter Hamiltonian (BSH) reads as:
\(\hat{H}_X = \hat{H}_0 + \hat{W} + \hat{J}\).
The first term $\hat{H}_0=\mathbbm{1}_{\mathrm{N}} \otimes \hat{H}_e - \hat{H}_h \otimes \mathbbm{1}_{\mathrm{B}}$ 
corresponds to the free electron-hole pair energies. 
It manifests as an effective six-dimensional tight-binding Hamiltonian on the basis of localized hole-electron excitations $
    \mathcal{B}_X = \qty{ \ket{\mathrm{N}, \vb{p}}\otimes \ket{\mathrm{B}, \vb{n}} \  / \ \vb{p} \in \Lambda_\mathrm{N}, \vb{n} \in \Lambda_\mathrm{B}} $
whose elements $\ket{N, \vb{p}} \otimes \ket{B, \vb{n}}$ we shall write $\ket{\vb{p},\vb{n}}$ for short. Indeed, the latter can be interpreted as effective real-space orbitals indexed by three coordinates for the hole position ($\vb{p}$) and three coordinates for the electron position ($\vb{n}$). 
The second term $\hat{W}$ is the direct electron-hole interaction. In this basis, it can be approximated by $
    \mel{\vb{p},\vb{n}}{\hat{W}}{\vb{p}^\prime,\vb{n}^\prime} \approx \delta_{\vb{p}, \vb{p}^\prime} \delta_{\vb{n}, \vb{n}^\prime} W(\abs{\vb{n}-\vb{p}})$
where $W$ is the real space screened Coulomb potential,\cite{Galvani2016} which can itself be modeled macroscopically, e.g. through the Rytova-Keldsyh potential.\cite{Rytova1967, Keldysh1979}
Finally, the third term $\hat{J}$ is the electron-hole exchange interaction which is here neglected since we expect the direct interaction to be the main contribution to the Bethe-Salpeter Kernel. 
To exploit this Hamiltonian in a periodic framework, we construct for each Nitrogen-Boron pair $\qty(\vb{p},\vb{n})$  \emph{of the unit cell} a Bloch state out of the translationally equivalent localized hole-electron pair states:\cite{Paleari2018, Sponza2018}
$\ket{\vb{p},\vb{n},\vb{\Omega}}_{\vb{Q}} = \frac{1}{\sqrt{N}} \sum_{\vb{\rho} \in \mathcal{R}} e^{-i\vb{Q}\cdot\vb{\rho}} \ket{\vb{p}+\vb{\rho}, \vb{n}+\vb{\rho}+\vb{\Omega}}$
where the $\vb{\Omega}$ are elements of the system's Bravais lattice $\mathcal{R}$, $N$ is the number of unit cells and $\vb{Q}$ is the difference in momenta between electron and hole states.
In the case of one orbital per atom, we find:\cite{Paleari2018, Sponza2018}
\begin{multline}
\label{eq:W_Bloch}
    {}_{\vb{Q}}\!\mel{\vb{p},\vb{n},\vb{\Omega}}{\hat{W}}{\vb{p}^\prime,\vb{n}^\prime,\vb{\Omega}^\prime}_{\vb{Q}^\prime} 
    \\ = \delta_{\vb{Q}, \vb{Q}^\prime}\delta_{\vb{p}, \vb{p}^\prime}\delta_{\vb{n}, \vb{n}^\prime} \mel{\vb{p},\vb{n}+\vb{\Omega}}{\hat{W}}{\vb{p},\vb{n}+\vb{\Omega}}
\end{multline}
where the latter matrix elements are in the real-space $\mathcal{B}_X$ basis, and:
\begin{multline*}
    {}_{\vb{Q}}\!\mel{\vb{p},\vb{n},\vb{\Omega}}{\hat{H}_0}{\vb{p}^\prime,\vb{n}^\prime,\vb{\Omega}^\prime}_{\vb{Q}^\prime} 
    \\ = \delta_{\vb{Q}, \vb{Q}^\prime}\delta_{\vb{p}, \vb{p}^\prime} \mel{\mathrm{B},\vb{n}+\vb{\Omega}}{\hat{H}_e}{\mathrm{B},\vb{n}^\prime+\vb{\Omega}^\prime}
    \\ - \delta_{\vb{Q}, \vb{Q}^\prime}\delta_{\vb{n}, \vb{n}^\prime}e^{-i\vb{Q}\cdot\qty(\vb{\Omega}-\vb{\Omega}^\prime)} \mel{\mathrm{N}, \vb{p}}{\hat{H}_h}{\mathrm{N},\vb{p}^\prime+\vb{\Omega}-\vb{\Omega}^\prime}
\end{multline*} 
Noticing that in the state $\ket{\vb{p},\vb{n},\vb{\Omega}}_{\vb{Q}}$ the electron and the hole are separated by the vector $\vb{R}=\vb{n}+\vb{\Omega}-\vb{p}$,
we can introduce a cutoff $R_{cut}$ and truncate away from the Hilbert space of pairs states with $\abs{\vb{R}}>R_{cut}$. This is a good approximation as long as we consider excitons with Bohr radius $a_B \ll R_{cut}$ (in practice we treat $R_{cut}$ as a convergence parameter), and allows the dimension of $\hat{H}_X$ to scale as $O(N_{orb}{R_{cut}}^d)$ where $N_{orb}$ is the number of orbitals in the unit cell and $d$ the system's dimension. Details of this procedure may be found in Supp. Mat.
As expected, this BSH is block diagonal in $\vb{Q}$. Here, we will focus on the $\vb{Q}=\vb{0}$ block, which is the only one which contributes to direct absorption.\cite{Onida2002,Rohlfing2000}. These equations are similar to the ones derived by Merkel and Ortmann\cite{Merkel2024} in a Wannier functions setting, with our perturbative approximations substituting for the explicit determination of Wannier functions. This avoids the need to diagonalize the single particle Hamiltonian to proceed to the excitonic calculation, pushing further the efficiency of this kind of approaches.

Having obtained $\hat{H}_X$, it remains to extract the corresponding optical response.
We evaluate the absorption spectrum of the system by computing the imaginary part of its dielectric function, $\varepsilon_2$, which for light polarized along the direction $\vb{e}$ and incoming photon energy $E=\hbar\omega$ is given by:\cite{Attaccalite2018}
\begin{equation}
\label{eq:TB_epsilon2}
    \varepsilon_2^{\vb{e}\vb{e}}\qty(E) \propto \frac{1}{E^2}\mel{\emptyset}{\vb{e}\cdot\hat{\vb{P}}\delta(E\mathbbm{1} - \hat{H}_X)\vb{e}\cdot\hat{\vb{P}}}{\emptyset}
\end{equation}
where $\hat{\vb{P}}$ is the momentum operator in the two-body representation and $\ket{\emptyset}$ (filled Fermi sea) is the many body ground state. Consequently, the computation of $\varepsilon_2\qty(E)$ amounts to the evaluation of the mean value of $\delta(E\mathbbm{1} - \hat{H}_X)$ in the state $\vb{e}\cdot\hat{\vb{P}}\ket{\emptyset}$ which can be approximated according to 
\begin{equation*}
\hat{\vb{P}}\ket{\emptyset} \approx
 \sum_{\vb{p},\vb{n}, \vb{\Omega}}   \mel{\mathrm{B},\vb{n}+\vb{\Omega}}{\hat{H}_{el}}{\mathrm{N},\vb{p}} \qty(\vb{n}+\vb{\Omega}-\vb{p}) 
 \ket{\vb{p},\vb{n}, \vb{\Omega}}_{\vb{0}}
\end{equation*}
where again the $\vb{p}$ and $\vb{n}$ run over the unit cell atomic positions and the $\vb{\Omega}$ run over the system's Bravais lattice.\cite{Galvani2016, GalvaniThesis, Attaccalite2018}

Since $\hat{H}_X$ is sparse in the $\ket{\vb{p},\vb{n}, \vb{\Omega}}_{\vb{0}}$ basis, a number of linear scaling algorithms are known to be computationally efficient for such a task. Here we leverage the KPM\cite{Weisse2006, Fan2021, Merkel2024} (using its implementation in KWANT\cite{Groth2014}). Its cost scales linearly with the dimension of $\hat{H}_X$.\footnote{More precisely, the expected scaling is $O(DM)$, where $D$ is the number of nonzero elements in $\hat{H}_X$, and $M$ is number of KPM moments, which fixes the effective broadening of the calculation.\cite{Weisse2006, Fan2021}} Thanks to the real-space truncation described above, the KPM thus offers a scaling of $O(N_{orb}{R_{cut}}^d)$ to compute $\varepsilon_2\qty(E)$ once $\hat{H}_X$ is known.
Further details on the influence of $R_{cut}$ on the computation of $\varepsilon_2\qty(E)$ may be found in Supp. Mat.

To illustrate the methodology, we consider the case of a monolayer of hBN in the presence of an Anderson disorder, which could mimic spatial energy variations of onsite potential due to environmental effects affecting either a supported or embedded BN layer inside a device. We use a previously established $p_z$ nearest-neighbor tight-binding parametrization,\cite{Galvani2016} along with a diagonal Anderson disorder $\hat{\mathcal{W}} = \sum_{\mu\in\qty{B,N},\vb{n}}\epsilon_{\vb{n}}\dyad{\mu,\vb{n}}$ added to the electronic Hamiltonian $\hat{H}_{el}$. Here the $\epsilon_{\vb{n}}$ are independently and uniformly distributed in $\qty[-\dis,\dis]$ with $\dis$ the strength of the disorder. We note that, compared to the pristine case, this yields an additional term to the BSH, of the form $\hat{\mathcal{W}}_X = \sum_{\vb{p},\vb{n}}\qty(\epsilon_{\vb{n}}-\epsilon_{\vb{p}})\dyad{\vb{p},\vb{n}}$. We model the screened Coulomb interaction \(W(\abs{\vb{n}-\vb{p}})\) using a Rytova-Keldysh potential\cite{Rytova1967, Keldysh1979} with a fixed polarizability radius $r_0 = 10 \text{ \AA}$ for all values of $\dis$.\cite{Galvani2016} Further computational details may be found in Supp. Mat.

\begin{figure}[t]
    \centering
    \includegraphics[width=\columnwidth]{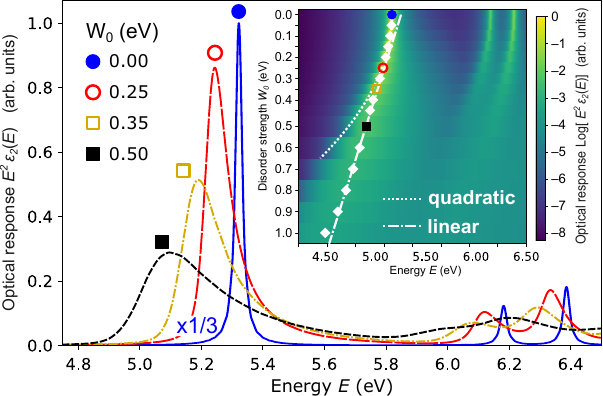}
    \caption{Main panel: Calculated optical response $E^{2}\varepsilon_{2}(E)$, normalized to $1/3$ of the main pristine excitonic peak, for different Anderson disorder strengths. Inset: color map of the same as a function of photon energy and disorder strength, superimposed with main peak position and fits of its disorder-induced shift.    }
    \label{fig:anderson}
\end{figure}

Figure \ref{fig:anderson} displays the evolution of the absorption spectrum as a function of disorder strength using the method described above.
We used $\sim 10^4$ atoms supercells and averaged over $200$ realizations of disorder for each value of $\dis$, for a total of $\sim 3000$ Bethe-Salpeter calculations. The inset shows results for all $\dis$s while the main panel displays selected linecuts.
We choose a Lorentz kernel for the KPM, yielding an approximately constant \(\eta_0 \approx 12.5\) meV lorentzian broadening of delta peaks,\cite{Weisse2006} as can be observed in the \(\dis = 0 \text{ eV}\) case. Beyond this case, we observe a disorder-induced non-symmetric broadening of the excitonic peaks, as well as a redshift of the peak maximum. 
This redshift can be fitted to a dependance in \(\dis^2\) for low values of the disorder, while it appears to be linear in \(\dis\) for higher disorder values (Fig.~\ref{fig:anderson} inset).
To quantify the disorder-induced broadening, we calculate the \emph{excess} Full Width at Half Maximum (FWHM) at given disorder, defined as the deviation of the measured FWHM from the zero disorder FWHM, $2\eta_0 \approx 25 \text{ meV}$.
 We found that it is well approximated by \(\dis^2A\) where \(A\approx 1.1 \text{ eV}^{-1}\),
at least for disorder strengths up to \(\dis \sim 0.5 \text{ eV}\).
This overall behavior of the main absorption peak is reminiscent of exciton-phonon lineshapes, with the disorder-induced redshift being the analogue of the polaronic shift\cite{Lengers2020,Toyozawa1958} as well as the situation of Frenkel excitons in molecular aggregates,\cite{Fidder1991, Dominguez2004} (as discussed in Supp. Mat.).

\begin{figure}
    \centering
    \includegraphics[width = 0.99\linewidth]{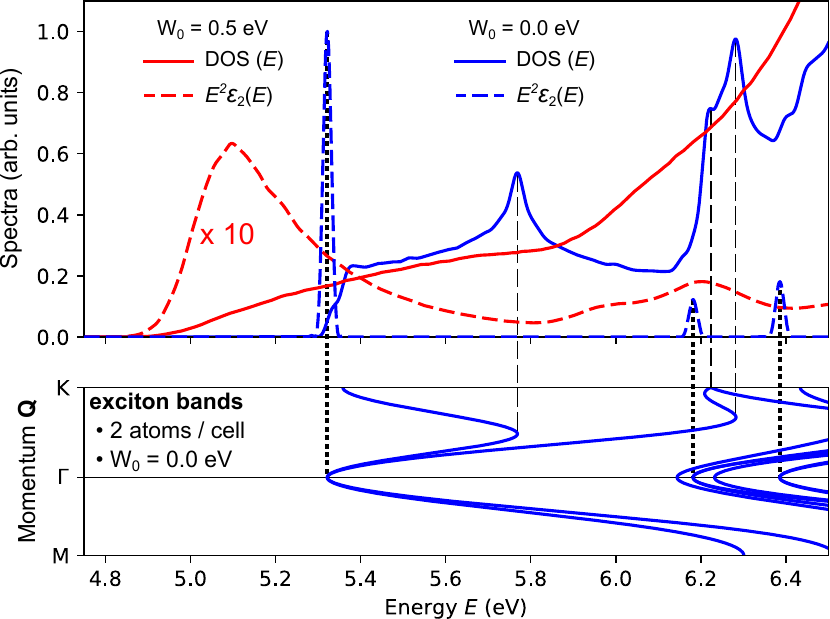}
    \caption{Top panel: exciton DOS (solid line) and absorption spectrum (dotted line) for hBN single layer with no (blue) and \(\dis=0.5 \text{ eV}\) (red) Anderson disorder, computed in a $10^5$-atom cells. Lower panel: exciton band structure of hBN single clean layer computed in the 2-atom unit cell via diagonalization. Across the two panels thin dashed lines highlight the correspondence between extrema of the band structure and van Hove singularities of the DOS while thick dotted ones link bright exciton states to peaks of $E^2\varepsilon_2(E)$.
    }
    \label{fig:dos}
\end{figure}

To better understand these behaviors and further validate the methodology, we compute the excitonic density of states i.e. the density of states of the BSH for monolayer hBN in the pristine \(\dis=0\text{ eV}\) and moderately disordered \(\dis=0.5\text{ eV}\) cases, as shown in Fig.~\ref{fig:dos}. These densities of states have been computed using the KPM as \(\rho(E) \propto \text{Tr}\qty[\delta\qty(E\mathbbm{1} - \hat{H}_X)]\) with the stochastic trace approximation.\cite{Fan2021} 
Indeed, the excitonic DOS is analogous to an optical response with unitary oscillator strengths. Namely, if~\eqref{eq:TB_epsilon2} can be cast in the form $E^2\varepsilon_2(E) \propto \sum_\Psi \abs{\mel{\emptyset}{\vb{e}\cdot\hat{\vb{P}}}{\Psi}}^2\delta(E-E_\Psi) $ where the \(\Psi\) run over the eigenstates of \(\hat{H}_X\) and the \(E_\Psi\)s are the corresponding eigenenergies, then $\rho(E) \propto \sum_\Psi \delta(E-E_\Psi)$. 

The pristine case serves here as a validation benchmark. In fact, any supercell calculations at $\dis=0$~eV must predict a spectrum consistent with the excitonic band structure of the unitary 2-atom cell. This is indeed what we found, as one can appreciate by comparing the top panel of Fig.~\ref{fig:dos} (KPM spectrum on a $10^{5}$-atom supercell) with the bottom panel (excitonic band structure on the 2-atom unitary cell via diagonalization).
We find an excellent match between the energies of the bright states in the band structure and the optical peaks in the KPM calculation.

In the disordered case, Fig.~\ref{fig:dos} shows that the excitonic DOS is smeared, and the optical response is due to a wide continuum of optically active states, in contrast with the pristine case where only discrete states from the $\vb{Q}=\vb{0}$ sector of the 2-atom primitive unit cell had nonvanishing oscillator strengths. This situation arises from the breaking of the translational invariance, which allows the mixing of states from different \(\vb{Q}\) sectors of the primitive unit cell (now all folded into the $\vb{Q}=\vb{0}$ sector of the supercell).
Because the Anderson disorder is diagonal, it does not affect the two-body momentum operator \(\hat{\vb{P}}\). The eigenstates of the disordered system thus acquire their optical activity through the component that the pristine ($\dis =0$) bright states have in the mixing. Keeping this in mind, we can explain qualitatively (i) why the peak broadens asymmetrically and (ii) why it redshifts.
(i) Since bright pristine states lie at the bottom of their excitonic bands, the states that can hybridize with them under the action of disorder are situated at higher energies (towards the blue). In contrast, no pristine states are available towards the red. This suggests a lineshape that broadens asymmetrically, more strongly towards the blue, as we observe numerically. (ii) At the perturbative level, we also expect that the lowest bound bright state, being at the bottom of its band, will undergo level repulsion from the states above, and therefore shift towards the red.
A further confirmation of this interpretation comes from  applying  second order perturbation theory to the single band case (see Supp. Mat.). 
Indeed, this approach predicts a redshift of the pristine peak proportional to $\dis^2$ for small values of $\dis$, in agreement with our numerical observations up to $\dis \sim 0.15$~eV (see inset of Fig.~\ref{fig:anderson}), after which an affine scaling is observed instead.

We now focus on the spatial characteristics of the exciton upon disorder through the real-space analysis of the excitonic eigenfunctions.
Albeit at a higher cost than the extraction of spectral responses, this is also possible through our method. Indeed, our construction provides computationally efficient access to a sparse real-space BSH with dimension 
\(O(N_{orb})\). It is therefore amenable to sparse diagonalization routines,\cite{lehoucq1998arpack} which allow the efficient extraction of a few eigenstates close to an energy of interest, which itself can be determined through a linear scaling computation of the DOS or \(\varepsilon_2\).

\begin{figure}
    \includegraphics[width = 0.99\linewidth]{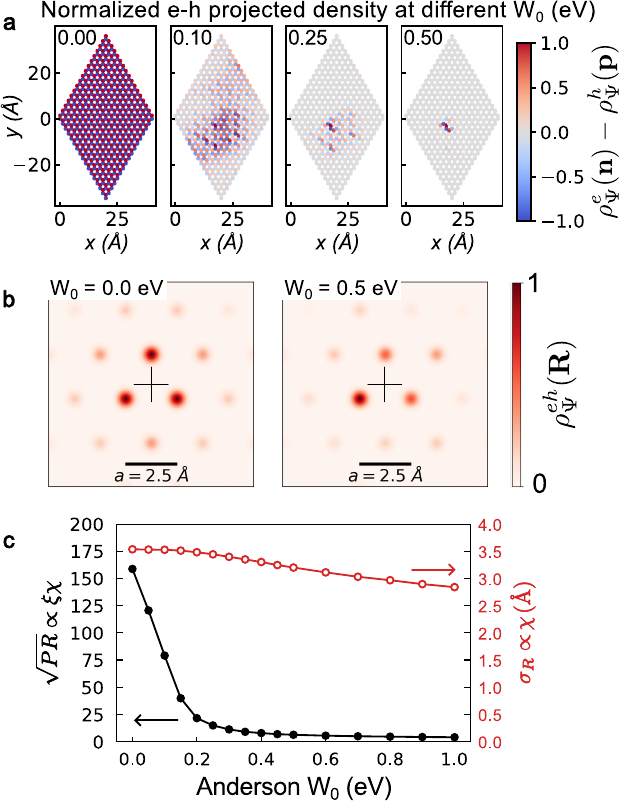}
    \caption{\textbf{a}: Normalized electron - hole projected density (see text) for the lowest bound excitonic eigensubspace with increasing disorder, illustrating localization of the center of mass coordinate.
    \textbf{b}: Relative coordinate probability density $\rho^{eh}(R)$ for the lowest bound exciton at $\dis=0$ and $\dis=0.5$ eV (crosses mark the origin).
    \textbf{c}: Disorder dependence of the localization $\sqrt{PR}$ and compactness ($\sigma_{R}$), averaged over the 40 lowest bound excitonic states in a $\sim 10^{4}$-atom supercell.}
    \label{fig:SL_WF}
\end{figure}

Using this approach, we extract, for several values of disorder strength \(\dis\), the lowest bound excitonic states in the system. These states are naturally obtained under the form \(\ket{\Psi} = \sum_{\vb{p},\vb{n},\vb{\Omega}}\Psi_{\vb{p},\vb{n},\vb{\Omega}} \ket{\vb{p},\vb{n},\vb{\Omega}}\), whose high dimensionality makes their direct representation impractical in large unit cells. For this reason, we rely on effective single coordinate local densities of states (LDOS - see Supp. Mat.). To visualize the center of mass degree of freedom of a given state \(\ket{\Psi}\), we use effective LDOSs for holes, \(\rho_\Psi^h (\vb{p}) = \sum_{\vb{n},\vb{\Omega}} \abs{\bra{\vb{p},\vb{n},\vb{\Omega}}\ket{\Psi}}^2\), and for electrons, \(\rho_\Psi^e (\vb{n}) = \sum_{\vb{p},\vb{\Omega}} \abs{\bra{\vb{p},\vb{n},\vb{\Omega}}\ket{\Psi}}^2\). For the sake of clarity, Fig.~\ref{fig:SL_WF}a depicts \(\rho_\Psi^e (\vb{n})-\rho_\Psi^h (\vb{p})\) for the lowest bound excitonic eigenspace in a supercell made of only 578 atoms.\cite{Naik2022,Li2024_imaging}  Here, one disorder realization $\{ \epsilon_\mathbf{n} \}_{\mathbf{n}}$ was fixed and only scaled up with \(\dis\) for consistency. While at \(\dis = 0\) we recover the expected Bloch states delocalized over the entire supercell, we observe very clearly the localization of the center of mass with increasing disorder strength. 

We can likewise examine the localization along the relative coordinate, \(\vb{R} = \vb{n}+\vb{\Omega}-\vb{p}\), that is the compactness of the exciton. Contrary to the center of mass localization, which is absent in the pristine case and enabled by disorder, a localization along \(\vb{R}\) is already induced by the direct interaction term even in the pristine case. In Fig.~\ref{fig:SL_WF}b, we depict for the lowest bound exciton an effective relative coordinate density $\rho_\Psi^{eh}\qty(\vb{R}) = \sum^\prime_{\vb{p}, \vb{n}, \vb{\Omega}} \abs{\bra{\vb{p},\vb{n},\vb{\Omega}}\ket{\Psi}}^2$ where the prime constrains the sum to those coordinates satisfying $\vb{n}+\vb{\Omega}-\vb{p} = \vb{R}$, and we took care to average the probabilities for degenerate states at $\dis = 0 \text{ eV}$ (as in Fig.~\ref{fig:SL_WF}a). \cite{Galvani2016,UriaAlvarez2024,Ferreira2019_hBNSL,Attaccalite2018}
This quantity approximates the probability density for the electron-hole separation and it reduces to the familiar fixed-hole representation in perfect crystals.\cite{Galvani2016,UriaAlvarez2024}  In the presence of disorder, we observe that despite a breaking of the triangular \(C_3\) symmetry at the single wavefunction level, the overall picture is not strongly modified. The electron-hole pair remains tightly bound, consistently with the strong $E_b \approx -1.9 \text{ eV}$ binding energy in the pristine case,\cite{Galvani2016,Wirtz2006_BNDimensionality} and well within the cutoff \(R_{cut} = 20 \text{ \AA}\) used for this calculation. 

A more quantitative way to evidence this is by introducing two characteristic quantities, namely the compactness $    \sigma_{R}\qty(\ket{\Psi}) = \sqrt{\sum_{\vb{R}}\rho_\Psi^{eh}\qty(\vb{R}){\vb{R}}^2-\left[\sum_{\vb{R}}\rho_\Psi^{eh}\qty(\vb{R}){\vb{R}}\right]^2}$ and the localisation  $PR\qty(\ket{\Psi}) = \left. 1 \middle/ \sum_{\vb{p}, \vb{n}, \vb{\Omega}}\abs{\bra{\vb{p},\vb{n},\vb{\Omega}}\ket{\Psi}}^4 \right.$. 
The first is the standard deviation of $\vb{R}$ which can be interpreted as an effective Bohr radius. For example $a_{\text{eff}} \approx \sqrt{2}\sigma_{R}$ for 1s states.
The latter is a 2-coordinate participation ratio. In the continuum limit, a trial probability density $\abs{\Psi\qty(\vb{p}, \vb{n})}^2\propto e^{-\abs{\vb{L}}/\xi}e^{-\abs{\vb{R}}/\chi}$, which is localized both in its center of mass $\vb{L}=\qty(\vb{n}+\vb{p})/2$ and relative  coordinate $\vb{R}=\vb{n}-\vb{p}$, has $PR\qty(\ket{\Psi}) \propto \xi^2 \chi^2/a^4$, where $a = 2.5 \text{ \AA}$ is the pristine unit cell parameter.
Since we have seen that $\chi \sim \sigma_{\vb{R}}$ is roughly constant with disorder, $\sqrt{PR\qty(\ket{\Psi})}$ behaves as a semi-quantitative indicator of the center of mass localization. In Fig.~\ref{fig:SL_WF}c, we plot $\sigma_R$ and $\sqrt{PR}$ averaged over the 40 lowest bound states in a $10^4$-atom supercell as a function of $\dis$.
In this supercell, we find that
for $\dis >0.1$~eV, this quantity is well fitted by an expression of the form $\sqrt{PR\qty(\ket{\Psi})} \approx A/\dis^2 + B$. The 0.1~eV threshold appears as a consequence of the finiteness of our computational
supercell (here $10^4$ atoms) because when $\dis \rightarrow 0$, $\xi \rightarrow \infty$. This analysis demonstrates that low-lying excitons in Anderson disordered hBN conserve the high compactness they have in the perfect monolayer ($\sqrt{2}\sigma_R \approx 5$~\AA) while drastically losing mobility as a consequence of a strong $\sim \dis^2$ Anderson localization.

In conclusion, we have established a linear-scaling route to excitonic spectroscopy in very large hBN-derived disordered models by mapping the Bethe--Salpeter problem onto a sparse real-space Hamiltonian of localized electron--hole pairs and evaluating optical spectra without diagonalization. The method exposes a regime inaccessible to conventional small-cell treatments: disorder does not merely broaden excitonic lines, but activates dark spectral weight, drives a quadratic-to-linear redshift crossover, and localizes the exciton center of mass while preserving the compact relative electron--hole structure. This real-space perspective opens a path toward exciton engineering in materials where broken translation symmetry is intrinsic, including electrically controlled and Hall-active excitons in transition-metal dichalcogenides (TMD),\cite{Onga2017,Ciarrocchi2019,Jauregui2019} long-lived interlayer excitons and moiré exciton lattices in TMD heterostructures,\cite{Rivera2015,Montblanch2021,Alexeev2024,Zhang2020_TwistMoirExc} and close-to-($30^\circ$) BN/BN quasicrystals whose large approximants and self-similar local registries require genuine scaling analysis.\cite{Roman-Taboada2023,Sponza2024,Latil2023,Roux2025} The same methodology may also be extended to exciton diffusion through further linear-scaling methods,\cite{Fan2021} a particularly relevant direction given that diffusion in current hBN samples has been observed to be limited by static disorder rather than phonons.\cite{Roux2024}

\section*{Acknowledgments}

This work was inspired by the insightful discussions and guidance of François Ducastelle, our mentor and friend, who left us far too soon. The ICN2 is funded by the CERCA programme / Generalitat de Catalunya and supported by the Severo Ochoa Centres of Excellence programme, Grant CEX2021-001214-S, funded by MCIU/AEI/10.13039.501100011033. This work was performed with funding from PN: Proyecto PID2022-138283NB-I00 (MCIN/ AEI /10.13039/501100011033) and by `FEDER Una manera de hacer Europa', as well as `Proyecto PCI2021-122092-2 financiado por MCIN/AEI /10.13039/501100011033 y por la Unión Europea NextGenerationEU/PRTR.' This work has also benefitted from the French National Research Agency grant number ANR-21-CE09-0016 (EXCIPLINT project).

\bibliography{biblio}

\end{document}